# Detection of a Satellite of the Trojan Asteroid (3548) Eurybates – A Lucy Mission Target

K. S. Noll[1,2], M. E. Brown[3], H. A. Weaver[4], W. M. Grundy[5], S. B. Porter[6], M. W. Buie[6], H. F. Levison[6], C. Olkin[6], J. R. Spencer[6], S. Marchi[6], T. S. Statler[7]

**Abstract**

We describe the discovery of a satellite of the Trojan asteroid (3548) Eurybates in images obtained with the Hubble Space Telescope. The satellite was detected on three separate epochs, two in September 2018 and one in January 2020. The satellite has a brightness in all three epochs consistent with an effective diameter of $d_2$ =1.2±0.4 km. The projected separation from Eurybates was $s$ ~1700-2300 km and varied in position, consistent with a large range of possible orbits. Eurybates is a target of the *Lucy* Discovery mission and the early detection of a satellite provides an opportunity for a significant expansion of the scientific return from this encounter.

_______________________
[1] Corresponding author: keith.s.noll@nasa.gov
[2] Goddard Space Flight Center, Code 693.0 Greenbelt MD 20771
[3] Division of Geological and Planetary Sciences, Caltech, Pasadena CA 91125
[4] Johns Hopkins Applied Physics Laboratory, Laurel MD 20707
[5] Lowell Observatory, Flagstaff AZ 86001
[6] Southwest Research Institute, Boulder CO 80302
[7] NASA Headquarters, Washington DC 20546

1. **Introduction**

Starting in 2018 the first deep satellite search of the Trojan asteroids that are targets of NASA's *Lucy* mission (Levison et al. 2017) was conducted using the Hubble Space Telescope (HST; Noll et al. 2018). Deep searches with high contrast are required to detect small satellites and observations with HST are uniquely well-suited for this purpose. Here we report the discovery of a previously unknown satellite of the Lucy target Eurybates, a $d$=63.9±0.3 km body (Grav et al. 2012) that is the largest member of the only known major disruptive collisional family in the Trojans (Brož & Rozehnal 2011). From these observations, we have identified a roughly 1-km-diameter satellite with the temporary designation S/2018 (3548) 1.

The detection of a satellite of Eurybates is not a completely unexpected result. Satellites and binaries occur in most small body populations (e.g. Margot et al. 2015, Noll et al. 2020). Three Trojans have had directly identified satellites or binary companions (Merline et al. 2001, Marchis et al. 2006, Noll et al. 2016) with additional binary candidates identified by lightcurves (Mann et al. 2007, Sonnett et al. 2015, Ryan et al. 2017). The reported yield of direct searches has been low, with only 2 of 94 resulting in detections (Marchis et al. 2006, Merline et al. 2007, Noll et al. 2016). However, the large size range of targets searched (20 km < $d$ < 200 km), differing instruments, filters and exposure times, and many unreported observational details make a formal understanding of the expected frequency of satellites from these data impossible. Other analog populations, such as compositionally similar objects found in the Outer Main Belt (see Discussion) lead to lower limits for the frequency of small satellites on the order of 5%. A small satellite around a primary the size of Eurybates would most likely be formed by collision (Margot et al. 2015), consistent with what is known about the Eurybates collisional family.

2. **Observations and Analysis**

The first two sets of observations of Eurybates occurred on 12 and 14 September 2018 (Table 1) using HST's Wide Field Camera 3 (WFC3) as part of HST general observer (GO) program GO-15144. Images were obtained with the F555W filter (WFPC2 V, $\lambda_{eff}$ = 530.8 nm, width = 156.2 nm) in the UVIS2-M1K1C-SUB subarray. A sequence of four 30s exposures was followed by four 350s and then by a final four 30s exposures. Each group of four was dithered using the standard WFC3-UVIS-DITHER-BOX pattern. The final group of four was offset by x, y = 0.172, 0.148 arcsec. Post-flash was set to 12 for the 30s exposures and 5 for the 350s exposures to minimize charge transfer efficiency losses during readout. A possible satellite was identified by visual inspection in images on both dates (Fig. 1). In each of the two combined images shown, four individual 350s images from the visit were registered and median-combined to remove cosmic ray artifacts and to increase S/N. (For the observations with HST a "visit" is a single sequence of observations executed in one or more orbits using the same pair of guide stars.) The images show that the satellite moved by 1.7±1.1 pixels (see below) in the two days between these observations, consistent with a bound satellite. The satellite is detectable in individual frames with a consistent brightness and PSF, ruling out the possibility of an artifact in the combined frame from coincident cosmic rays. Both parallax within a visit and the heliocentric orbital motion of Eurybates between visits rule out background sources.

Table 1
Observational Circumstances: Eurybates and S/2018 (3548) 1

| Date | Time* (UT) | Orient† (°) | Filter | $n_{exp}$ | $t_{int}$ (sec) | R (AU) | D (AU) |
|---|---|---|---|---|---|---|---|
| 09/12/18 | 09:26 | -70.305 | F555W | 4 | 30 | 5.371 | 4.599 |
|  | 09:45 |  |  | 4 | 350 |  |  |
|  | 10:05 |  |  | 4 | 30 |  |  |
| 09/14/18 | 09:07 | -70.306 | F555W | 4 | 30 | 5.370 | 4.619 |
|  | 09:26 |  |  | 4 | 350 |  |  |
|  | 09:48 |  |  | 4 | 30 |  |  |
| 12/11/19 | 06:16 | -62.996 | F350LP | 2 | 30 | 5.074 | 5.122 |
|  | 06:37 |  |  | 6 | 330 |  |  |
| 12/21/19 | 22:03 | -61.862 | F350LP | 2 | 30 | 5.067 | 5.280 |
|  | 22:35 |  |  | 6 | 330 |  |  |
| 01/03/20 | 07:11 | -64.498 | F350LP | 2 | 30 | 5.059 | 5.452 |
|  | 07:31 |  |  | 6 | 330 |  |  |

*Mid-time of the observations. †ORIENTAT WFC3 header keyword value for distortion-corrected files gives position of detector y axis in degrees East of North

Based on these initial observations, we sought and obtained three additional orbits to recover and confirm the satellite. Observations were obtained as part of GO-16056 in December 2019 and January 2020. For these observations the wider F350LP filter ( Long pass, $\lambda_{eff}$ = 584.6 nm, width = 475.8 nm) and the UVIS2-C512C-SUB subarray were used. Two 30s exposures were followed by three pairs of 330s exposures. Each pair of observations was dithered with the WFC3-UVIS-DITHER-LINE pattern. For the second and third pair of long exposures, the pattern was offset by x, y = 0.092, 0.098 and 0.185, 0.197 arcsec respectively. A post-flash of 12 was used for the 30s exposures; no post-flash was required for the longer exposures. The satellite was detected again in images obtained on 03 January

2020, confirming its existence (Fig. 2). It was not detected in images obtained on 11 and 21 December 2019; we ascribe this to the satellite having been at a smaller angular separation from Eurybates where the background from the primary was too high for the faint satellite to be detected which occurs at a radius of ~ 0.4 arcsec from Eurybates for an object with the brightness of the observed satellite.

The images taken on 12 and 14 September 2018 and 03 January 2020 show the satellite at separations of 0.569±0.015, 0.511±0.019, and 0.581±0.023 arcsec respectively (Table 2). Positions were measured in each of the individual distortion corrected files (except for one of the four images from 14 September 2018 where a nearby cosmic ray makes a position determination very uncertain). The uncertainty of the *x*- and *y*-positions in the WFC3 instrument frame was estimated to be ±0.25 pixel for Eurybates and ±0.5 pixel for the satellite in the images taken in September 2018. For the 2019-2020 observations, the uncertainty was greater, ±0.5 pixel for Eurybates (because of saturation) and ±1 pixel for the satellite. Uncertainties were propagated to the separation values shown in Table 2 and are in good agreement with the variance in the individual position determinations. The radial separation and position angle of the satellite, relative to Eurybates, were similarly calculated from the measured positions and yield projected distances of $s$ = 1903±50, 1716±64, and 2303±91 km.

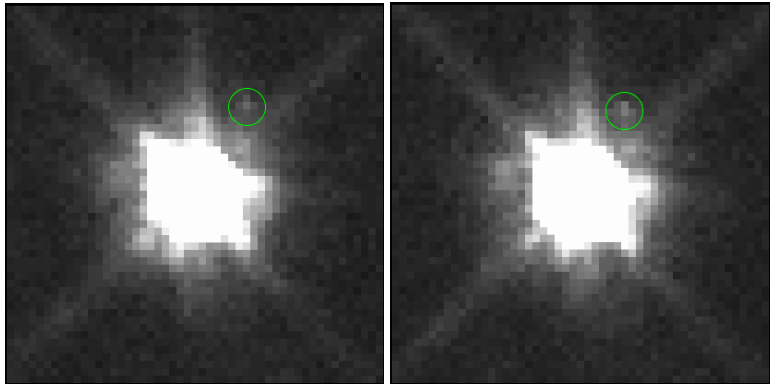

**Fig. 1** A 2×2 arcsec portion of HST WFC3 images of Eurybates from 12 and 14 September 2018 (left to right) taken with the F555W filter. Each image consists of four registered, flat-fielded, geometrically uncorrected exposures that have been median-combined. The images are shown using a linear stretch from -20 to 200 e⁻. The satellite is circled in each image. Other image features are due to the point-spread-function of Eurybates.

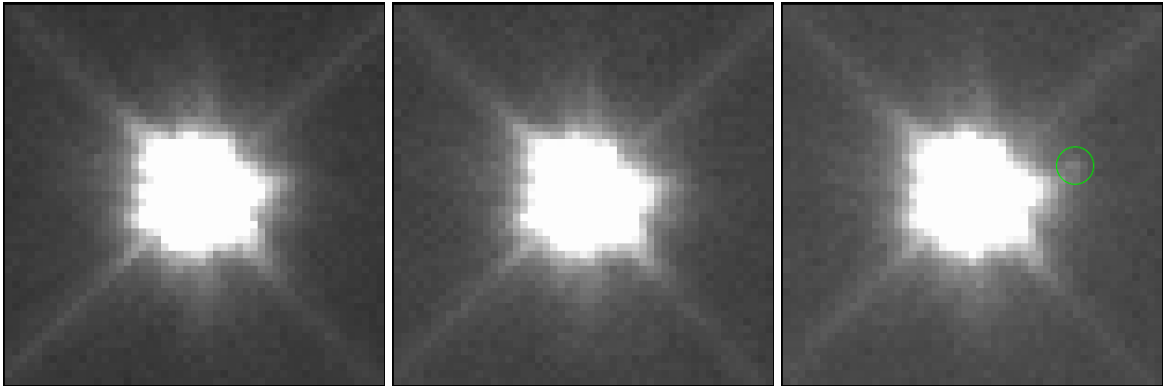

**Fig. 2** A 2×2 arcsec portion of HST WFC3 images of Eurybates from 11 and 21 December 2019 and 03 January 2020 (left to right) taken with the F350LP filter. Each image consists of six registered, flat-fielded, geometrically uncorrected exposures that have been median-combined. The images are shown using a linear stretch from -44 to 440 e$^-$. The stretch covers the same range of brightness as figure 1 by accounting for differences in filter throughput and the different observational circumstances (on 03 January 2020) compared to September 2018. The satellite is not detected in the images from 11 and 21 December 2019; it is circled in the 03 January 2020 image.

We measured the satellite's brightness in each of the flat-fielded images using a 3×3 pixel box centered on the brightest pixel. We chose the smallest possible aperture centered on the brightest pixel in order to minimize noise from the highly variable background. The background was determined from a similar 3×3 aperture at a mirror position relative to Eurybates to take advantage of the inherent symmetry of the PSF when comparing the azimuthally-varying background. The variance in the background aperture was assumed to also apply to the source aperture and was used as the uncertainty for the source aperture. The background-subtracted counts were averaged to yield the final result for each visit. For photometric uncertainties we chose the larger of the fully propagated uncertainty or the variance in counts from individual exposures within a visit. On the 14th of September we were only able to use three of the four 350s exposures, one was unusable for brightness measurements because of a nearly coincident cosmic ray. Because of the larger uncertainty for this visit, we report the relative brightness on that date as an estimate only. In all cases, the uncertainty of the brightness measurement for the satellite dominates the total uncertainty. We similarly measured the brightness of Eurybates in the 30s exposures from each visit (to avoid saturation) and determined the relative brightness of the satellite from the ratio of the two brightness measurements. Using $d_1$ = 63.9±0.3 km for the diameter of Eurybates (Grav et al. 2012), the brightness ratio of the secondary relative to the primary, and assuming the satellite has the same albedo as Eurybates, we derive an effective diameter of $d_2$ = 1.2±0.4 km for the satellite. Eurybates has an absolute magnitude of H = 9.55±0.30 (Veres et al. 2015); applying same brightness ratio yields H(satellite) = 18.25$\pm^{0.8}_{0.6}$.

**Table 2**
**Observed Properties: Eurybates and S/2018 (3548) 1**

| GO | visit | satellite relative position | | | | s† | Δ |
|---|---|---|---|---|---|---|---|
| | | x (mas) | y (mas) | r (mas) | ϕ(°)* | (km) | (mag) |
| 15144 | 21 | 279±11 | 485±11 | 569±15 | 259.8±1.6 | 1903±50 | $8.7\pm^{0.5}_{0.3}$ |
| 15144 | 22 | 237±14 | 453±14 | 511±19 | 262.1±2.2 | 1716±64 | *8.9* |
| 16056 | 01 | - | - | <400 | - | | - |
| 16056 | 02 | - | - | <400 | - | | - |
| 16056 | 03 | 550±18 | 186±18 | 581±23 | 224.2±2.3 | 2303±91 | $8.7\pm^{0.6}_{0.5}$ |

Estimated value shown in italics; * degrees East of North; † projected separation

### 3. Discussion

Existing positional constraints (Table 2, Fig. 3), including non-detections, can be used to determine whether physically reasonable orbits are possible. Indeed, we find a wide range of possible orbits for assumed bulk densities ranging from $\rho = 500 – 2500$ kg m$^{-3}$. However, it has not been possible with existing positional information to constrain period, eccentricity, semimajor axis or any other orbital parameters.

One possible way to further constrain the orbit that we considered was to estimate the timescale for tidal circularization (Goldreich and Soter, 1966). For small satellites where $m_2 \ll m_1$, this timescale is given by

$$\tau_{circ} = 3.437\times10^5 \, Q_2 \, \rho_1^{-3/2} \, \rho_2 \, d_1^{-9/2} \, d_2^{-2} \, a^{13/2} \quad (1)$$

where the leading numerical constant is in SI units and gives $\tau_{circ}$ in seconds when the remaining dimensional quantities are also in SI units. The shortest possible self-consistent tidal-circularization timescale can be computed by assuming an object on an already circularized orbit with the smallest semimajor axis consistent with observations, $a=2212$ km (the 1-σ-low value for the projected separation observed on 03 January 2020), and by adopting the smallest possible values for $Q_2$ and $\rho_2$ and the largest possible values for $d_1$, $d_2$, and $\rho_1$. We assign a primary diameter of $d_1 = 63.9$ km (Grav et al. 2012) and an assumed bulk density of $\rho_1 = 2500$ kg m$^{-3}$, a satellite diameter of $d_2 = 1.6$ km and density $\rho_2 = 500$ kg m$^{-3}$, and a dimensionless tidal quality factor $Q_2 = 10$ (Goldreich and Soter, 1966). With these choices we find a tidal circularization time, $\tau_{circ} = 7$ Gyr. If we choose nominal values instead, $a=2303$ km, $\rho_1 = 1500$ kg m$^{-3}$, and $d_2 = 1.2$ km, we find $\tau_{circ} = 34$ Gyr. The circularization timescale is most sensitive to the value of the semimajor axis that is assumed. Given the ≤4.5 Gyr age of the system, we conclude that the eccentricity of the mutual orbit cannot be constrained to be $e \approx 0$ from *a priori* considerations alone.

The detection of a satellite of Eurybates makes it the fourth Trojan with a directly observed companion. (617) Patroclus and (16974) Iphthime are roughly equal mass binaries (Merline et al. 2001, Noll et al. 2016) and (624) Hektor is a bilobed primary with a small secondary satellite (Marchis et al. 2006). Several more Trojans have been identified as possible close or contact binaries from their lightcurves (Mann et al. 2007, Sonnett et al. 2015, Ryan et al. 2017). Of these, the Eurybates system is most similar to Hektor, but Eurybates' satellite is smaller both in absolute and relative terms.

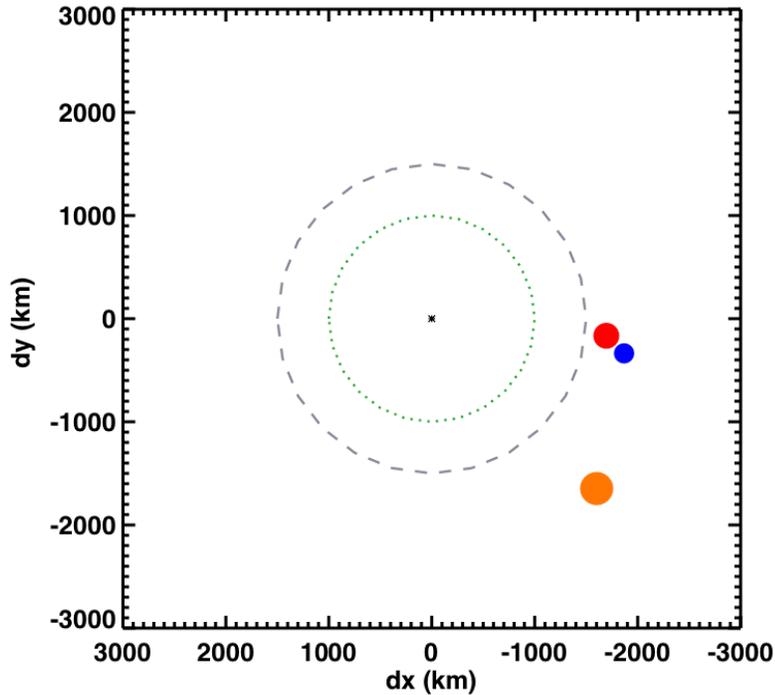

**Fig. 3** The separations of the satellite relative to Eurybates (black) are shown in a frame of reference with North up and East to the left: 12 September 2018 (blue), 14 September 2018 (red), and 03 January 2020 (orange). Symbol sizes are equal to the 3σ positional uncertainty. The dashed gray circle is the approximate boundary inside of which the satellite is too faint relative to the brightness of Eurybates to be detected and represents the positional uncertainty of non-detections on 11 and 21 December 2019. The green dotted circle has a radius of 1000 km, the planned *Lucy* close-approach distance during encounter.

We can broaden the basis for comparison by also considering all known small satellites (defined here, somewhat arbitrarily, as systems where $d_2/d_1 < 0.1$) in the Main Belt, where there are twelve known systems that meet this criterion. Within this larger group, the relative size of Eurybates' satellite, $d_2/d_1 \sim 0.019$, still stands out as small (Table 3). This may be a function of observational biases – only a limited number of asteroids have had similarly deep searches for satellites. Peneius, the satellite of the C-type Main Belt asteroid (41) Daphne, has $d_2 < 2$ km and orbits a primary with $d_{1eff}$ = 174 km, giving it a size ratio of $d_2/d_1 < 0.011$ (Conrad et al 2008). The smaller of the two known satellites of (130) Elektra, S/2014 (130) 1, also has a smaller relative size with $d_2/d_1 \approx 0.01$. As shown by Fig. 4, satellites appear to cluster at separation of less than 0.04 $R_{Hill}$, where $R_{Hill}$ is the Hill radius. Both Peneius and S/2014 (130) 1 orbit more than 3× closer to their primaries in $a/R_{Hill}$ than the Eurybates satellite. Only Huenna has a satellite at a greater separation, $a/R_{Hill}$ = 0.218 (Margot 2003). S/2018 (3548) 1 occupies a previously empty portion of this phase space.

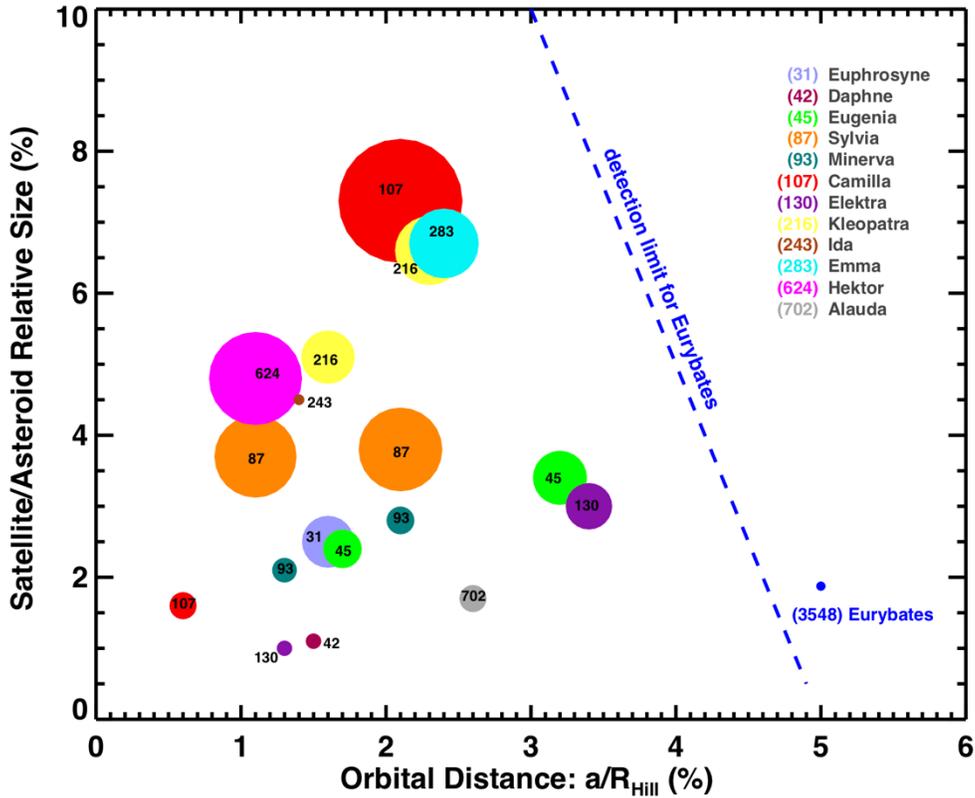

**Fig. 4** The orbital distance of satellites relative to the Hill radius, $a/R_{Hill}$, and relative sizes of satellites, $d_2/d_1$, are plotted. Symbol sizes are proportional to the satellite diameter, $d_2$. The small relative and absolute size of Eurybates' satellite is notable and occurs just within the HST detection limit for Eurybates (blue dashed). The detection limit shown is the separation where the PSF background is greater than the signal from a source of a given relative size. This limit differs for every object and is a function of the size of the system mass, the distance of the system from Earth, the depth and resolution of the image. Eurybates is one of the smallest and most distant primaries for which a satellite has been detected from an Earth-based telescope.

Among more distant small-body populations the number of identified small satellites is subject to observational limitations. In the Kuiper Belt, Pluto's small satellites have diameter ratios that range from $0.018 \geq d_2/d_1 \geq 0.0025$ (Weaver et al., 2005, Showalter et al. 2011, 2012). Many objects in the Kuiper Belt are known to be binary (Noll et al. 2020), but small satellites are too faint to detect for all but the largest transneptunian objects.

It is noteworthy that ten of the twelve Main Belt systems with small satellites are objects with C-complex (C,B,D,F,G,T) and P spectral types, i.e. those most similar to Eurybates and the Trojans in general. There are 193 Main Belt and Outer Main Belt asteroids in the same size range as the objects listed in Table 3 (84km ≤ d ≤ 300km) in these spectral classes, yielding a lower limit satellite frequency of $f_{sat} \geq 5.2\%$ (subject to unquantifiable uncertainties in spectral classification). This suggests that additional small satellites orbiting spectrally similar objects in the Trojans, Hildas, and Outer Main Belt might remain undetected due to observational limitations. It is unclear if the apparent affinity of satellites with C-complex and P spectral types is a function of composition that favors satellite formation or, simply, the product of observational biases. Six of the twelve Main Belt satellite systems are multiples, suggesting that the satellite formation mechanism is efficient at producing more than a single satellite. The targets of the *Lucy* mission are all C, P, and D spectral type objects so the much more sensitive search for satellites that will be possible during this mission will provide an important constraint on the overall abundance of small satellites.

**Table 3**
**Asteroids with Small Satellites ($d_2/d_1 < 0.1$)**

| Object | Spectral Type | $d_{1eff}$ (km) | $\Delta_{mag}$ (mag) | $d_2/d_1$ (%) | $a/R_{Hill}$ (%) | $a/r_1$ | $e$ | refs |
|---|---|---|---|---|---|---|---|---|
| (3548) Eurybates | C/X | 63.9±0.3 | $8.7\pm^{0.5}_{0.3}$ | 1.9±0.7 | >5 | >18 | ? | *this work* |
| (31) Euphrosyne | C | 267±2 | 8±0.8 | 2.5±0.9 | 1.6 | 5.1 | ? | [1] |
| (41) Daphne | C | 174±12 | *10* | <1.1 | 1.5 | *5* | ? | [2] |
| (45) Eugenia | FC | 206±6 | 6.1±0.1 | 3.4±0.1 | 3.2 | 11 | <0.01 | [3] |
|  |  |  | *8* | 2.4±0.5 | 1.7 | 5.9 | 0.11±0.02 | [4] |
| (87) Sylvia | P | 286±11 | 6.2±0.2 | 3.8±2 | 2.1 | 9.5 | <0.01 | [5] |
|  |  |  | *7.2* | 3.7±0.2 | 1.1 | 4.9 | 0.09±0.02 | [6] |
| (93) Minerva | C | 142±2 | *7.8* | 2.8±1.4 | 2.1 | 8.8 | <0.01 | [7] |
|  |  |  | *8.4* | 2.1±0.7 | 1.3 | 5.3 | 0.05±0.04 | [7] |
| (107) Camilla | C | 219±6 | 7.0±0.1 | 7.3±2.7 | 2.1 | 11 | <0.01 | [8] |
|  |  |  | 9.0±0.3 | 1.6±0.2 | 0.6 | 3 | ? | [9] |
| (130) Elektra | G | 197±2 | *7.6* | 3.0±0.7 | 3.4 | 13 | *0.08* | [10,11] |
|  |  |  | *9.9* | 1.0±0.7 | 1.3 | *5* | *0.16* | [11] |
| (216) Kleopatra | M | 135±6 | *5.9* | 6.6±1.2 | 2.3 | 10 | <0.01 | [12] |
|  |  |  | *6.5* | 5.1±1.2 | 1.6 | 6.7 | <0.01 | [12] |
| (243) Ida | S | 31.4±1.2 | *6.7* | *4.5* | 1.4 | 6.9 | >0.2 | [13] |
| (283) Emma | P | 135±2 | *5.9* | 6.7±3.7 | 2.4 | 8.6 | 0.12±0.01 | [14] |
| (379) Huenna | B | 87±2 | *5.9* | 6.6±1.4 | 21.8 | 76 | 0.222±0.006 | [15] |
| (642) Hektor | D | 184±10* | *6.6* | 4.8±1.3 | 1.1 | 10.4 | 0.31±0.03 | [16] |
| (702) Alauda | C | 202±5 | *8.8* | 1.7±0.4 | 2.6 | 12 | <0.01 | [17] |

Table values from compilation maintained by Johnston (2018) including derived and estimated values and other sources as listed. *Effective diameter of equal-volume sphere is listed for objects with known non-spherical shape. Numbers in italics are estimated values. Primary references listed: [1] Yang et al. (2020); [2] Conrad et al. 2008, [3] Merline et al. 1999, [4] Marchis et al. 2004, [5] Brown et al. 2001, [6] Marchis et al. 2005, [7] Marchis et al. 2009, [8] Storrs et al. 2010, [9] Marsset et al. 2016, [10] Merline et al. 2003a, [11] Yang et al. 2016, [12] Marchis et al. 2008 , [13] Belton et al 1996, [14] Merline et al. 2003b, [15] Margot 2003, [16] Marchis et al. 2006, [17] Rojo & Margot 2007.

Because the orbit of the Eurybates satellite is not yet constrained we can only place a lower limit on the semimajor axis, $a$. With a maximum observed separation of $s$ = 2303 km and no constraint on eccentricity we find $a$ > 1151 km where the limit is for the case where the satellite was observed at greatest elongation at apoapsis in an orbit with $e \approx 1$. Even this extreme assumption gives a wider separation relative to the radius of the primary and the Hill radius than all but one of the known Main Belt small satellite systems. Based on the high frequency of multiples and the available stable orbital space, it is possible that Eurybates could harbor one or more additional satellites orbiting closer to the primary, interior to the HST detection limit. It is unlikely, however, that any such satellites will be detected before the *Lucy* spacecraft arrives.

Another particularly interesting aspect of the existence of a satellite arises because Eurybates itself is the largest member of the only confirmed Trojan collisional family (Brož & Rozehnal 2011). The Eurybates collisional family consists of ~100 members with $d \geq 10$ km and potentially many more smaller members. It is natural to speculate that the collision that formed this family could also have resulted in one or more fragments surviving as bound satellites (Durda et al. 2004). Sylvia, a P-type asteroid in the Cybele region of the Outer Main Belt with two known satellites also has a collisional family (Vokrouhlický et al. 2010). These authors find that two other Cybele asteroids with small satellites, (107) Camilla and (121) Hermione, could have had their collisional families dynamically depleted, especially in some scenarios of Jupiter's late migration. Together, this is suggestive that collisions likely are responsible for the formation of small satellites in this population. The *Lucy* flyby of

Eurybates, therefore, presents an opportunity to study a likely collisional satellite at close range which will help constrain our understanding of this more generally applicable satellite-formation mechanism.

*Lucy* will fly by Eurybates in August 2027 at a distance of 1000 km, well within the Hill sphere, and closer than the projected distance of this satellite (Fig. 4). The discovery of this satellite now, before launch, means that it may be possible to determine the satellite orbit prior to encounter, thus increasing the scientific yield of the mission by enabling complementary observations of this satellite during the flyby.

**Acknowledgments**

This research is based on observations made with the NASA/ESA *Hubble Space Telescope* obtained from the Space Telescope Science Institute, which is operated by AURA under NASA contract NAS 5-26555. These observations are associated with programs 15144 and 16056.

**Note Added in Proof**

The Eurybates satellite was detected for a fourth time in observations with HST on July 19, 2020 and was not detected in observations obtained on August 3. The data are still insufficient to determine an orbit. Additional observations are planned.